# High-bandwidth photodetector enabled by frequency-domain equalization

Wenxin Deng[1,2,†], Hengsong Yue [1,2,†], Xiaoyan Liu[1,2], Jianhong Liang[3], Jianbin Fu[3], Shilong Pan[3], Tao Chu[1,2,*]
[1]College of Information Science and Electronic Engineering, Zhejiang University, Hangzhou 310027, China
[2]State Key Laboratory of Extreme Photonics and Instrumentation, Zhejiang University, Hangzhou 310027, China
[3]National Key Laboratory of Microwave Photonics, Nanjing University of Aeronautics and Astronautics, Nanjing 210016, China
[†]The authors contributed equally to this work.
* Corresponding author.
E-mail address: chutao@zju.edu.cn (T. Chu).

*Abstract*—**High-bandwidth germanium (Ge) photodetectors are crucial for silicon photonic integrated circuits. However, their bandwidth is restricted by carrier transit time and parasitic parameters. In this work, we propose an equalization photodetector (EqPD) utilizing the frequency response of a high-bandwidth photodetector $PD_A$ to subtract the frequency response of a low-bandwidth photodetector $PD_B$. With the response of $PD_B$ attenuating more severely than $PD_A$ at high frequency, the differential frequency response (the response of EqPD) can get higher values at high frequency than at low frequency, flattening the overall frequency response and expanding the bandwidth. Experimental results show that the EqPD has a bandwidth exceeding 110 GHz with a responsivity of 94 mA/W. A 100 Gbaud non-return-to-zero (NRZ) operation without digital signal processing is also demonstrated. To the best of our knowledge, this represents the highest bandwidth in a vertical Ge photodetector, providing a promising solution for high-speed photodetection in next-generation optical communication.**

## 1. Introduction

Artificial Intelligence and cloud computing drive increasing demand for processing a large amount of data, bringing a grand challenge of data communication [1], [2], [3]. To address the problem, silicon photonics offers a promising solution with the advantages of compatible fabrication processes with complementary metal–oxide–semiconductor, high integration density, low power consumption, and cost effectiveness [4], [5], [6]. Photodetectors are key elements in silicon photonic integrated systems for optical signal detection [7], [8], and germanium (Ge) is promising due to its fabrication compatibility and high absorption efficiency. Increasing requirements for high-speed data links create an urgent need for high-bandwidth Ge photodetectors (PDs). However, the bandwidth of Ge photodetectors is restricted by two main factors: carrier transit time and resistance and capacitance (*RC)* parasitic parameters [9].

A large amount of approaches have been proposed to improve the performance of Ge PDs. Some lateral PDs have been reported by narrowing the intrinsic Ge region to reduce the carrier transit time [10], [11], [12]. An impressive bandwidth of up to 265 GHz has been achieved through a fin-shaped lateral Ge PD and a narrow intrinsic Ge region of 100 nm [10]. However, this special process introduces a high dark current of 0.2 μA, and fabricating such a narrow intrinsic Ge is quite challenging. Vertical PDs are widely popular due to their compatibility with simple fabrication, avoiding mechanical polishing of Ge and silicon corrosion [13]. Several vertical Ge PDs with optimized *RC* parameters have also been reported by adjusting the Ge region size, growth, and silicon doping [14], [15], [16]. In addition, some studies utilize inductors to improve the bandwidth of Ge PDs to compensate for the junction-capacitance-induced roll-off [13], [17], [18], [19]. In Ref. [20], inductance and a U-shaped electrode are jointly used to optimize the parasitic parameters of a vertical Ge PD, and a bandwidth of 103 GHz with a dark current of 1.3 nA has been reported with an intrinsic Ge thickness of 350 nm. However, this method needs accurate inductance matching, and its bandwidth is still limited by the carrier transit time, which is finally determined by the fabrication processes. In conclusion, all these approaches have limitations, such as fabrication challenges, increased complexity, high dark current, and complex inductance parameters matching.

In this work, we propose and experimentally demonstrate an equalization PD (EqPD) which utilizes the frequency response of a relatively high-bandwidth $PD_A$ to subtract the frequency response of a low-bandwidth $PD_B$. As the frequency increases, the frequency response of $PD_B$ will attenuate more severely than $PD_A$, indicating that the differential frequency response (the ultimate frequency response of EqPD) can remain comparable to or exceed its low-frequency level at high frequencies. The EqPD can thereby flatten the overall frequency response and expand the bandwidth. The simulation results demonstrate that the EqPD can reach a bandwidth of up to 120 GHz, overcoming not only the calculated *RC*-limited bandwidth of 31 GHz but also the transit-time bandwidth of 80 GHz. Experimental results show that the bandwidth of the EqPD can be expanded to over 110 GHz with a responsivity of 94 mA/W. To the best of our knowledge, the EqPD has the highest bandwidth among vertical Ge PDs. An NRZ eye diagram of 100 Gbaud has also been demonstrated without digital signal processing (DSP). This study offers a novel solution for achieving ultrahigh-speed optical communication.

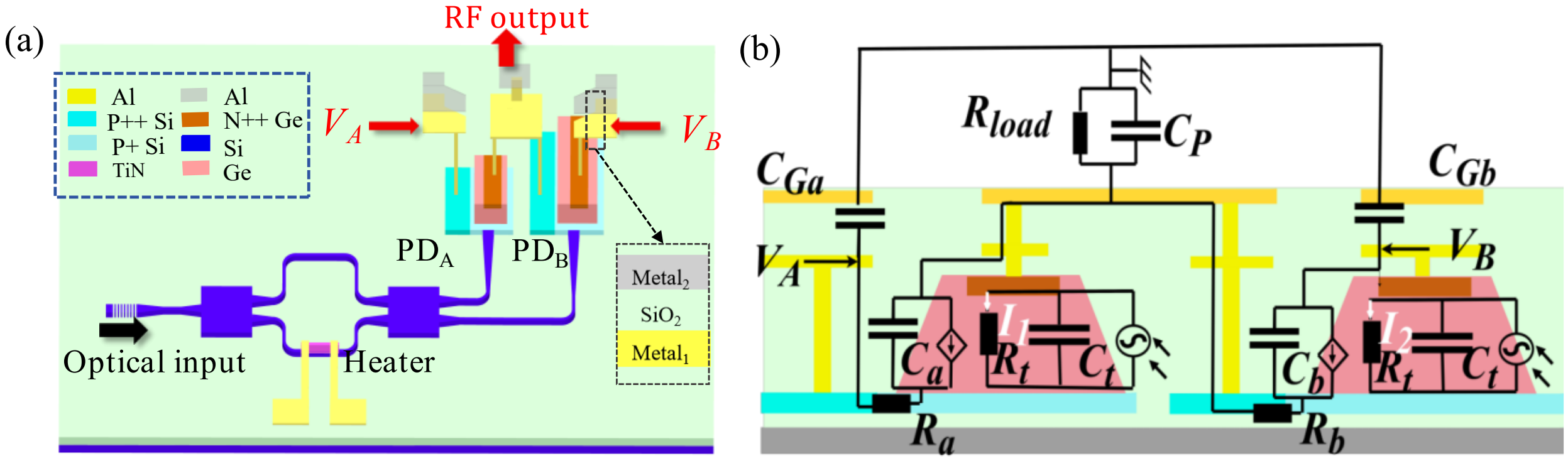

**Fig. 1.** (a) Three-dimensional diagram of the EqPD. (b) The cross-sectional schematic and equivalent circuit model of $PD_A$ and $PD_B$.

## 2. Principle and simulations

### *2.1. Principle*

A three-dimensional schematic diagram of the proposed EqPD is shown in Fig. 1(a). The EqPD consists of two differential PDs, $PD_A$ and $PD_B$, where the Ge area of $PD_A$ is smaller than that of $PD_B$, resulting in lower junction capacitance and higher bandwidth for $PD_A$. Their oppositely doped contact regions ( $N^{++}$ Ge for $PD_A$ and $P^+$ silicon for $PD_B$) share a common electrode, enabling current subtraction, where the RF output is determined by the frequency-dependent difference between $PD_A$ and $PD_B$. A thermo-tunable Mach–Zehnder interferometer (MZI) precedes the EqPD to controllably split the input optical power between $PD_A$ and $PD_B$, allowing adjustment of the optical power ratio of the two PDs.

In the electrical design, a dual-layer electrode architecture forms two metal–insulator–metal (MIM) capacitors between Metal1 and Metal2. The MIM capacitors can provide an AC return path for the RF signal while isolating the DC biasing from the output load effectively [21]. $V_A$ and $V_B$ are the bias voltages applied to the corresponding metal contact pads to ensure proper bias voltage of $PD_A$ and $PD_B$.

To clarify the operation of the EqPD, a cross-sectional view and equivalent circuit model are shown in Fig. 1(b) [21], where the photocurrents in Ge are modeled as current sources $I_1$ and $I_2$, yielding a differential output $I_{Eq} = I_1$-$I_2$. Generally, the frequency response of each PD is determined by both the carrier transit-time response and *RC* parasitic parameters, which can be approximately modeled as two cascaded low-pass transfer functions [19].

Firstly, the carrier transit-time response can be given by a first-order pole as:

$$H_t(f) = \frac{1}{1 + j2\pi f\tau_t},\ \tau_t = R_t C_t \qquad (1)$$

where $\tau_t$ represents the effective carrier-transit time constant, which physically describes the average time required for photogenerated carriers to traverse the Ge absorption region. In a vertical PD, it is mainly determined by the Ge thickness and carrier transit conditions. This carrier-transit response can be equivalently modeled as a first-order low-pass network consisting of resistance $R_t$ and capacitance $C_t$, as shown in Fig. 1(b). Since the $PD_A$ and $PD_B$ employ the same Ge thickness and similar carrier-transit conditions, their carrier-transit response can be approximated by a common term, denoted as $H_t(f)$. Therefore, the frequency response difference between PDA and PDB mainly originates from their distinct *RC* parasitic parameters.

The parameter *m* defines the fraction of the total photocurrent contributed by $PD_B$, while the complementary fraction (1-*m*) is contributed by $PD_A$. Based on this optical photocurrent distribution and the differential operation of the two PDs, the transfer function of the EqPD can be expressed as the weighted difference between two low-pass responses [19]:

$$H_{Eq}(f) = (1-m)H_a(f)H_t(f) - mH_b(f)H_t(f) \qquad (2)$$

$H_a(f)$ and $H_b(f)$ denote the transfer functions from the controlled current sources $I_1$ and $I_2$ of $PD_A$ and $PD_B$, respectively, to the load node, controlled by *RC* parasitic parameters and the mutual loading of the two PD branches. The *RC*-limited transfer function can be obtained directly from the equivalent circuit in Fig. 1(b).

For $PD_i$ ($i = a$ or $b$), the photocurrent source of the other $PD_k$ ($k \neq i$) is set to zero, while its electrical impedance remains connected to the output node. With $s = j2\pi f$, the effective load $Z_L$ and the impedance of branch $Z_k$ are:

$$Z_L(f) = \frac{R_{load}}{1 + sC_p R_{load}},$$

and

$$Z_k(f) = R_k + \frac{1}{sC_k} + \frac{1}{sC_{Gk}}.$$

Then, $H_a(f)$ and $H_b(f)$ can be described by the following equation:

$$H_i(f) = \frac{Z_k}{(1 + sC_p R_{load})(Z_L + Z_k)\left[1 + sC_i\left(R_i + \frac{1}{sC_{Gi}} + \frac{Z_L Z_k}{Z_L + Z_k}\right)\right]} \quad (3)$$

In the above formula, both index combinations $i=a$, $k=b$ and $i=b$, $k=a$ are applicable. Here, $C_a$, $C_b$, $R_a$, and $R_b$ denote the junction capacitances and resistances of $PD_A$ and $PD_B$; $C_{Ga}$, $C_{Gb}$, $C_p$, and $R_{load}$ account for the MIM coupling capacitors, parasitic capacitance, and load resistance. It should be noted that $C_{Ga}$ and $C_{Gb}$ are located in the bias-return branch rather than in series between the photogenerated current sources and $R_{load}$. The MIM capacitances interact with the synthesized parasitic elements in the complete small-signal network, thereby affecting the low-frequency response without producing the cutoff predicted by a simple $R_{load}C_{Ga}$ or $R_{load}C_{Gb}$ model. Formula (3) shows that $H_a(f)$ and $H_b(f)$ are monotonically decreasing as frequency increases.

*2.2 Simulation Results of Expanding Bandwidth*

The transfer function $H_{Eq}(f)$ can be adjusted by selecting the photocurrent distribution differences between $PD_A$ and $PD_B$ according to formula (2). The Ge thickness of both $PD_A$ and $PD_B$ is simulated as 0.5 µm, consisting of 0.05 µm thick $N^{++}$ Ge and 0.45 µm thick intrinsic Ge. The transit-time equivalent capacitance $C_t$ and resistance $R_t$ are estimated to be approximately 40 fF and 50 Ω, corresponding to a carrier transit-time limited bandwidth of 80 GHz. We set the Ge lengths of $PD_A$ and $PD_B$ to 8 µm and 17 µm respectively, and the width of both Ge to 6 µm. The sizes of $PD_A$ and $PD_B$ were selected to provide sufficient contrast in their frequency responses while preserving high responsivity for a single PD. The values of $C_a$ and $C_b$ are 17 fF and 42 fF respectively according to their sizes, while $R_a$ and $R_b$ are approximately 120 Ω and 160 Ω. $C_p$ and $R_{load}$ are set to 8 fF and 50 Ω, respectively. $C_{Ga}$ and $C_{Gb}$ are set to be 220 fF, which isolate the DC bias sources from the RF signal path while enabling independent biasing of the two PD branches. Since these capacitors are connected within the bias branches rather than directly in series with the RF output load, their influence is determined by the complete branch impedance.

Fig. 2(a) shows the calculated transfer function $H_{Eq}(f)$ for different values of $m$, where severe high-frequency attenuation is observed without equalization ($m = 0$). Increasing $m$ flattens the frequency response with increasing high-frequency peaking, leading to substantially reduced high-frequency attenuation over a broad bandwidth for $m \geq 0.3$.

The 3-dB bandwidth of EqPD can be obtained by solving the equation $|H_{Eq}(f_{3dB})|^2 = |H_{Eq}(0)|^2/2$. The normalized output RF response of the EqPD under various equalization conditions is illustrated in Fig. 2(b). All DC gains are standardized to 0 dB to enable clear visualization and quantitative comparison of the relative 3-dB bandwidth characteristics. When $m$ is zero, the calculated bandwidth is only 24 GHz, mainly restricted by RC parasitics. Furthermore, the simulated $S_{21}$ exhibits small fluctuations at low frequency, which are mainly caused by MIM capacitance, whose frequency-dependent impedance introduces slight variations in the circuit response.

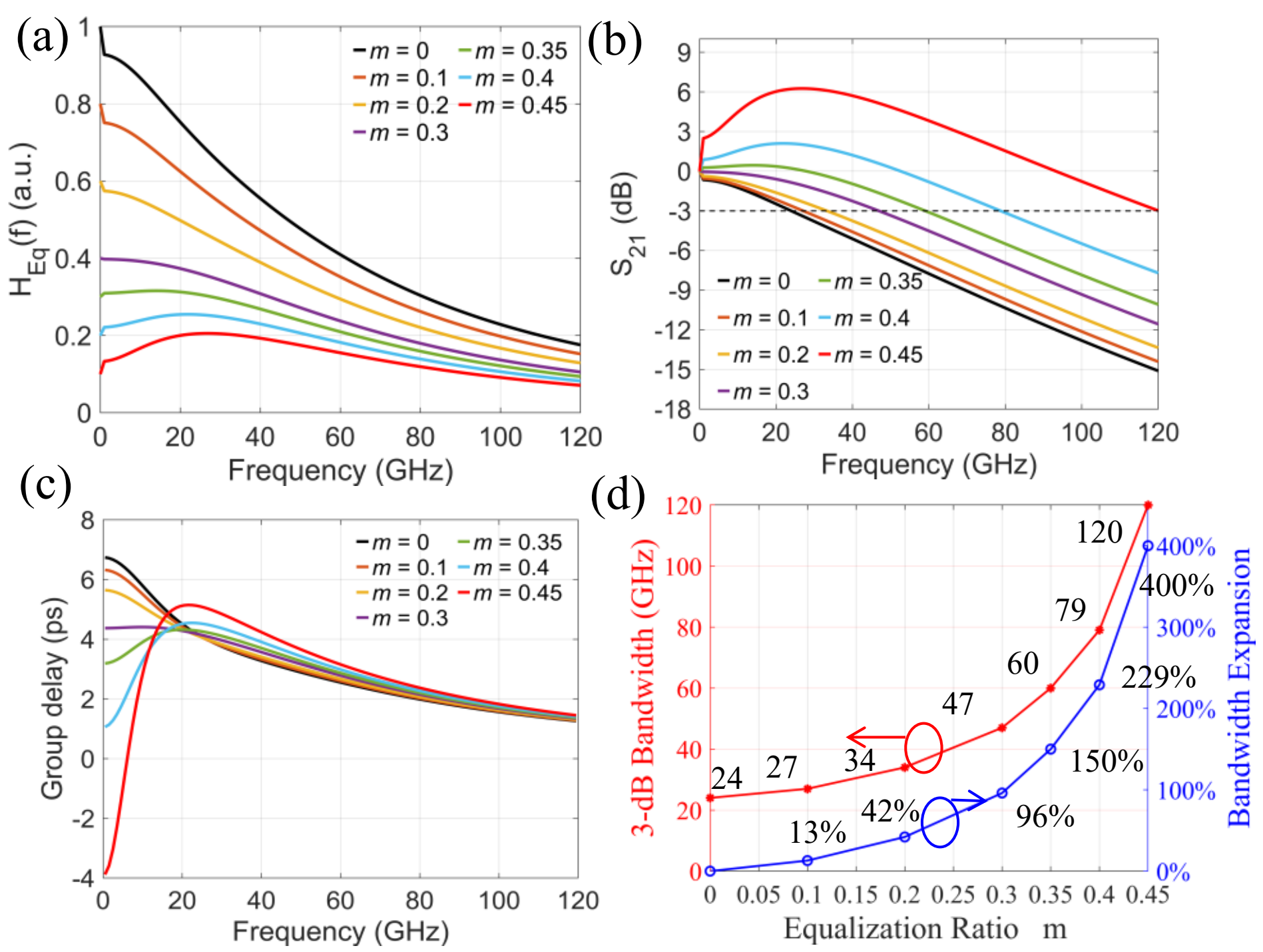

**Fig. 2.** (a) Simulated transfer function and (b) normalized $S_{21}$ of the EqPD with and without equalization. All DC gains are standardized to 0 dB. (c) The simulated group delay with different $m$. (d) The 3-dB bandwidth and bandwidth expansion of EqPD with different $m$. The parameter $m$ is the photocurrent fraction from $PD_B$.

The results also demonstrate enhanced frequency-response peaking and improvement in relative 3-dB bandwidth characteristics with increasing values of $m$. When $m = 0.1$, 0.2, 0.3, and 0.35, the corresponding simulated 3-dB bandwidths are 27, 34, 47, and 60 GHz, respectively. It is worth noting that the carrier transit-time-limited bandwidth, fundamentally governed by the intrinsic Ge thickness, establishes an inherent physical boundary for the 3-dB bandwidth of a single PD. Theoretically, if the minimum thickness of the intrinsic Ge is determined by the fabrication process, the maximum bandwidth is critically limited by the value [22]. The $RC$ parasitic parameters play a similar role in limiting bandwidth. However, when $m$ = 0.4 and 0.45, the simulated 3-dB bandwidth enhancement reaches 229% and 400%, corresponding to 79 GHz and 120 GHz, respectively. It is worth noting that the $RC$-limited bandwidths of $PD_A$ and $PD_B$ are 31 and 18 GHz, while their carrier transit-time-limited bandwidths are both 80 GHz due to identical intrinsic-Ge thickness. This means the bandwidth of EqPD can extend the effective 3-dB bandwidth beyond the bandwidth limitation of an individual PD branch determined by RC parasitics and carrier-transit response, which is of great significance.

Fig. 2(c) shows the simulated group delay with different $m$. As $m$ increases, the group delay exhibits a shift in the low-frequency region (< 20 GHz). When $m$ increases from 0 to 0.3, the change in group delay gradually decreases at low frequencies (< 20 GHz), with only a 0.05 ps change at $m$ = 0.3. As $m$ further increases, the change in group delay at low frequencies also increases. When $m$ is 0.45, the group delay shows a sharper transition of about 8 ps. However, across all photocurrent ratios $m$ (from 0 to 0.45), the group delay converges significantly at high frequencies (20-120 GHz), maintaining a smooth and relatively flat variation of about 3 ps. This indicates that the EqPD maintains a smooth group-delay response over 20–120 GHz, limiting additional phase distortion in the spectral region that contributes to high-speed signal transitions.

Fig. 2(d) summarizes the simulated 3-dB bandwidth and the bandwidth expansion of the EqPD at different values of $m$. The bandwidth increases significantly with increasing $m$. In bandwidth-limited high-speed links, frequency-dependent attenuation, particularly the excessive suppression of high-frequency components, aggravates intersymbol interference (ISI), thereby degrading signal integrity. By properly adjusting the photocurrent fraction ratio $m$, the EqPD can compensate for this frequency-dependent attenuation and achieve a more balanced frequency response, resulting in improved waveform quality for high-speed signal transmission.

## 3. Fabrication and measurements

### *3.1 Fabrication*

The device was fabricated on a silicon-on-insulator wafer with a 220-nm-thick silicon top layer and a 2-μm-thick buried oxide layer. The manufacturing processes and services were provided by Advanced Micro Foundry in Singapore. A 500 nm Ge layer was epitaxially grown on a $P^+$-doped silicon platform, with the top 50 nm Ge region heavily doped to an $N^{++}$ concentration. The doping concentrations were controlled to achieve a level of ~$10^{20}$ $cm^{-3}$ for $P^{++}$-Si and ~$10^{19}$ $cm^{-3}$ for $P^+$-Si. The structural design parameters of $PD_A$ and $PD_B$ were consistent with the simulation design specifications. Fig. 3 presents the microscopic image of the fabricated EqPD structure. Optical power was coupled to the chip through a grating coupler and then coupled into the MZI. Experimentally, $m$ is tuned by varying the optical power splitting ratio of the MZI. DC voltages for $PD_A$ and $PD_B$ are provided through the "$V_A$" and "$V_B$" electrodes, and the RF signal is extracted from the "S" electrode pad.

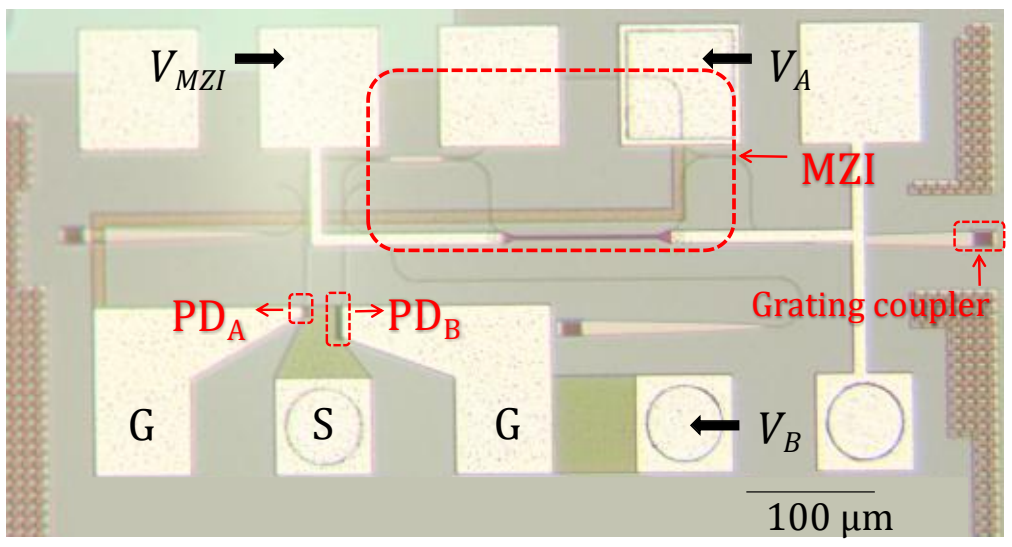

**Fig. 3.** Microscope image of the fabricated EqPD.

### *3.2 Measurement*

*3.2.1. Dark Current:* Figs. 4(a) and 4(b) show the dark current characteristics of the EqPD using a voltage source (Keysight B2901A). Fig. 4(a) shows the dark-current of the EqPD as the bias voltage of $PD_A$ changes from -3 to 0 V. $V_B$ was fixed at +1 V. Because $V_B$ is connected to the $N^{++}$-Ge top contact of PD, this polarity reverse-biases the junction and promotes depletion of the intrinsic Ge. Positive values in Figs. 4(a) and 4(b) indicate that the dark current of $PD_A$ is lower than that of $PD_B$.

Fig. 4(b) presents the differential dark current near the 10 pA level. The differential dark current reaches the 10 pA range under specific bias conditions, which mainly originates from the cancellation of the DC common-mode component between the two PDs. However, it should be noted that this DC dark current suppression cannot lead to a reduction in noise, since shot noise and thermal noise associated with the individual PDs cannot be canceled through the differential configuration.

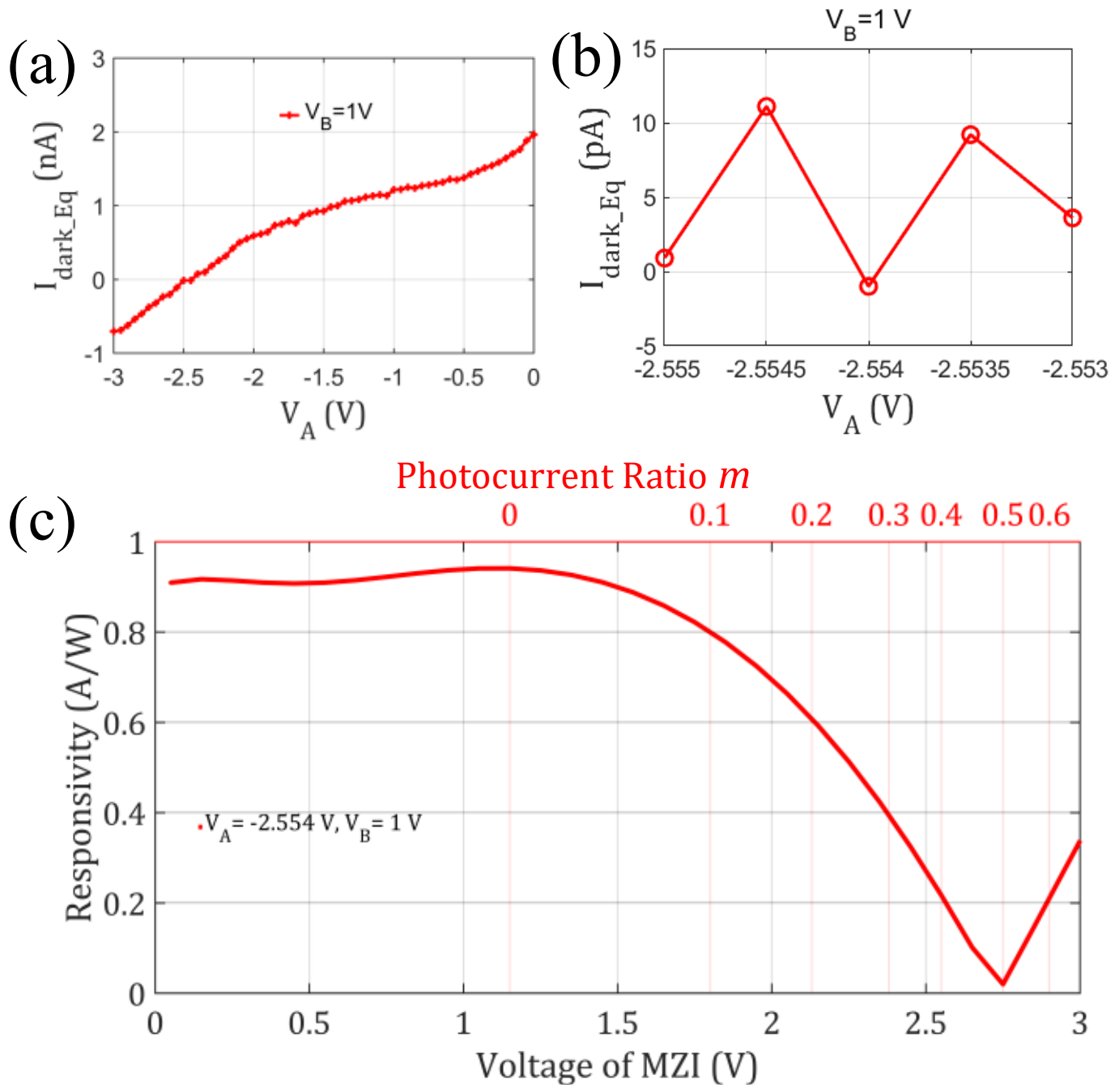

**Fig. 4.** (a) The measured dark current of the EqPD ($I_{dark_Eq}$) with different bias voltages of $PD_A$ when $V_B$ is 1 V. (b) The measured dark current of the EqPD near the 10 pA level. (c) The measured responsivity of the EqPD at a wavelength 1550 nm as a function of the MZI bias voltage.

*3.2.2. Responsivity:* Fig. 4(c) shows the responsivity and photocurrent ratio *m* of EqPD at 1550 nm as a function of the MZI bias voltage. The input optical power during testing was 5 dBm, with an overall link loss of 5.7 dB subtracted. When $PD_A$ is biased at -2.554 V and $PD_B$ at 1 V, the responsivity of EqPD at $m = 0$ is 0.94 A/W. The selected bias condition enables differential suppression of the DC current offset between the two PD branches, but AC noise contributions from each PD are not canceled. As the bias voltage of MZI changes, the optical powers delivered to $PD_A$ and $PD_B$ change, thus the responsivity of EqPD shows a periodic trend of decreasing and increasing. The maximum responsivity of EqPD is 0.94 A/W ($m = 0$), and the minimum responsivity is close to 0, which indicates that the photocurrent of $PD_A$ and $PD_B$ is nearly canceled. The responsivity decreases approximately linearly with increasing *m* ($0 < m < 0.5$). The relationship between *m* and the voltage of MZI can be obtained from Fig. 4(c).

*3.2.3. 3-dB Bandwidth:* The small-signal characteristics of the EqPD are measured as shown in Fig. 5(a) using a vector network analyzer (Keysight N5245B) and a 110-GHz lightwave component analyzer (Newkey GOCA-110) [23]. During high-frequency measurements, a commercial 110-GHz GSG RF probe is used, while DC biases ($V_A$, $V_B$, and $V_{MZI}$) are applied via gold wires to an external printed circuit board (PCB). The RF ground pads on $Metal_2$ are capacitively coupled to $Metal_1$ through a thin $SiO_2$ layer, forming the integrated coupling capacitors $C_{Ga}$ and $C_{Gb}$, which provide a low-impedance RF return path while isolating the DC bias. The photocurrent ratio *m* ranges from 0 to 0.472. Fig. 5(a) shows that as m increases, the low-frequency $S_{21}$ magnitude gradually decreases, and the frequency curve becomes flatter. The frequency responses for $m = 0.1$ and $m = 0.2$ slightly exceed those of $m = 0$ at high frequency, which may arise from the phase difference in the bias application path between $PD_A$ and $PD_B$. The normalized $S_{21}$ parameters are displayed in Fig. 5(b) to compare the 3-dB bandwidth.

The 3-dB bandwidths of the EqPD without and with equalization ($m$ =0, 0.1, 0.2, 0.3, and 0.35) are 17, 25, 33, 55, and 65 GHz respectively. When $m = 0.4$, the EqPD keeps low attenuation compared to low-frequency within the 60 GHz frequency range and demonstrates a bandwidth of up to 73 GHz. At $m = 0.45$, the normalized response at 110 GHz is only 0.53 dB below its low-frequency level; therefore, the 3-dB bandwidth lies beyond the 110-GHz measurement range. The effective responsivity is 0.094 A/W.

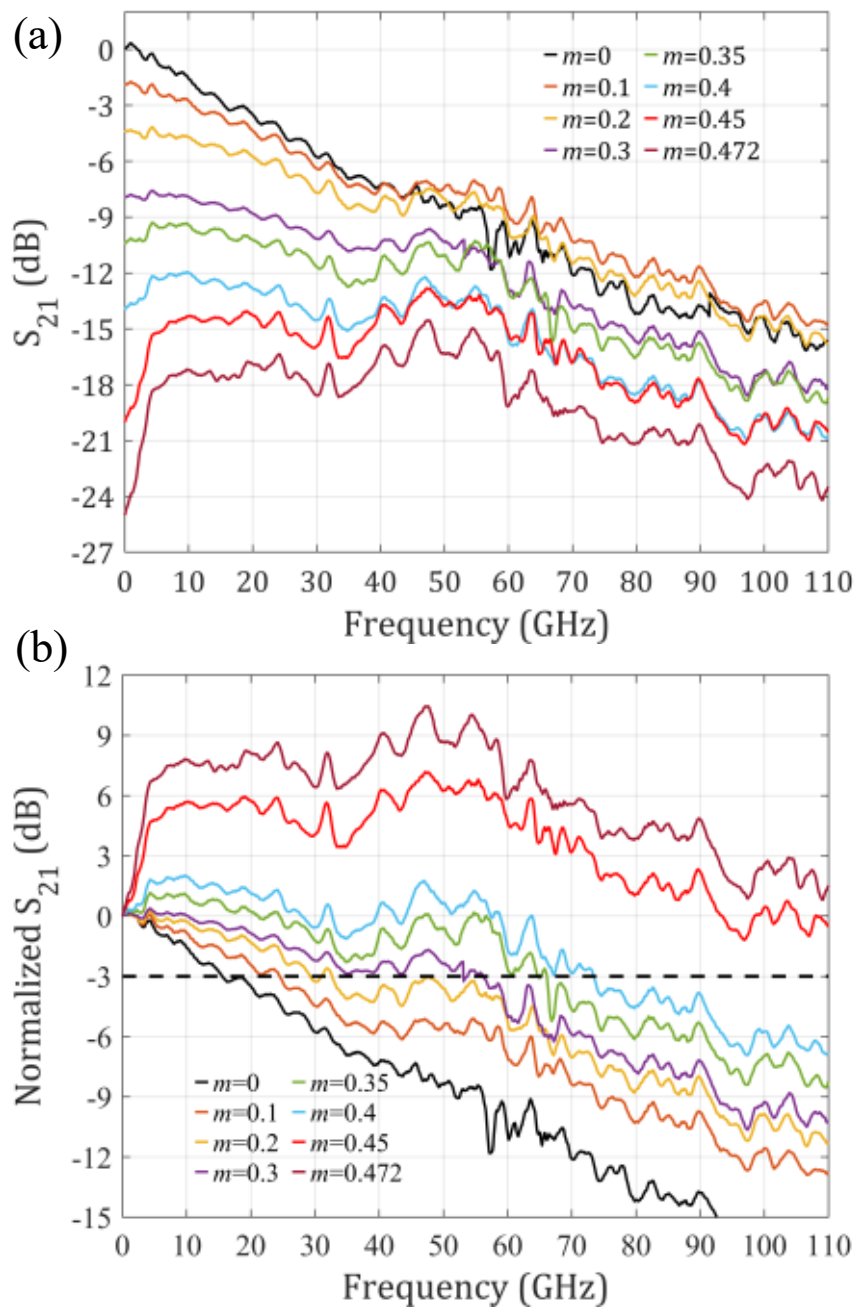

**Fig. 5.** The measured (a) $S_{21}$ and (b) normalized $S_{21}$ of the EqPD at $V_A$ = -2.554 V and $V_B$ = 1 V.

As $m$ increases toward 0.5, the bandwidth can be further widened. When $m$ =0.472, there is no loss in the frequency response relative to low frequency within the 110 GHz range.

*3.2.4. Eye Diagrams:* To characterize the high-speed performance of the proposed PD, we tested the eye diagrams of the proposed PD. Fig. 6(a) shows the experimental setup for NRZ eye diagram measurement. The multiplexer (SHF C603B) accepted two serial data streams from the bit pattern generator (SHF 12104A), resulting in a high-baud-rate electrical NRZ signal. The signal was amplified by a 60 GHz amplifier (SHF S804B). The output eye diagrams and signal-to-noise ratio (SNR) were recorded using a digital serial analyzer (Tektronix DSA8300).

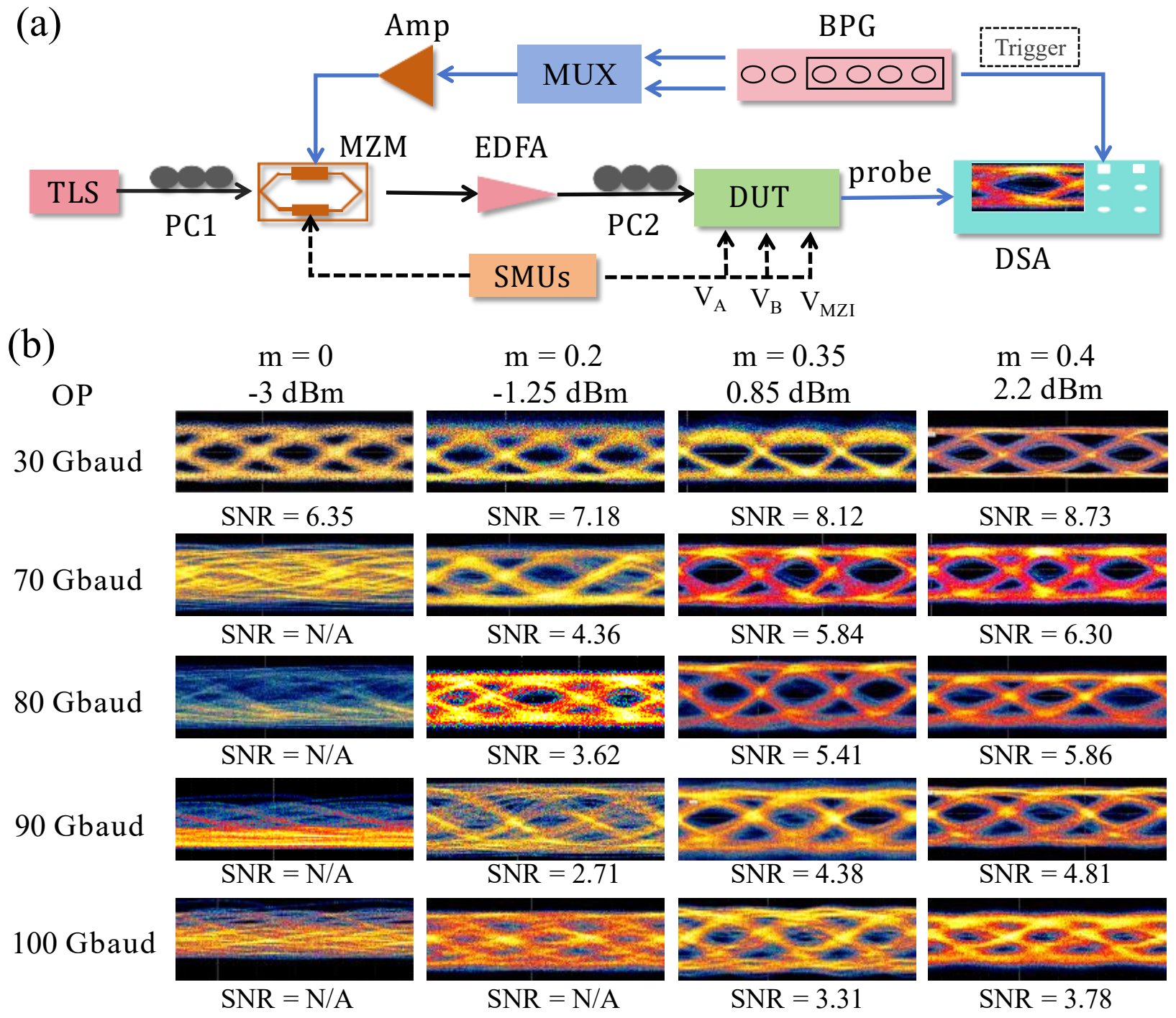

**Fig. 6.** (a) Experimental setup for NRZ eye diagram measurement. The black dashed lines and blue lines represent DC and AC electrical signals, respectively. The black solid lines represent optical signals. TLS: tunable laser source; PC, polarization controller; BPG, bit pattern generator; MUX, multiplexer; Amp, microwave amplifier; MZM, Mach-Zehnder modulator; EDFA, erbium-doped fiber amplifier; SMU, source-measure unit; DUT, device under test; DSA, digital serial analyzer. (b) Eye diagrams at different baud rates and $m$ values of the EqPD. OP, the optical power at the EqPD input after deducting the fiber and grating-coupler losses. $V_A$ = -2.554 V and $V_B$ = 1 V. N/A indicates that the SNR could not be reliably extracted because of eye closure.

**Table 1. Performance comparison of Ge PDs.**

| Ref. | Type | Bandwidth-enhancement approach | O/E Frequency Response | Responsivity (mA/W) |
|---|---|---|---|---|
| [12] | Lateral | Poly-Si+ thin Ge | -3 dB at 67 GHz | 740 |
| [24] | | ultra-thin Ge | -3 dB at 100 GHz | 1050 |
| [10] | | Narrow intrinsic Ge | -3 dB at 265 GHz | 300 |
| [19] | Vertical | *RC* parasitic Optimization+inductance | -3 dB at 80 GHz | 890 |
| [25] | | Light-trapping | -3 dB at 67 GHz | 1050 |
| [20] | | *RC* parasitic Optimization+inductance | -3 dB at 103 GHz | 950 |
| [13] | | *RC* parasitic Optimization+inductance | -3 dB at 75 GHz | 810 |
| This Work | | Differential frequency-domain equalization | -0.53 dB at 110 GHz | 94 |

In the experiment, the input optical power was correspondingly increased to compensate for the responsivity reduction caused by differential subtraction. This adjustment compensates for the signal amplitude for different equalization conditions, while the bandwidth enhancement originates from the frequency-domain reshaping of the EqPD response rather than optical power compensation. Fig. 6(b) compares the measured eye diagrams under different symbol rates and various $m$ values. When $m$ =0, 0.2, 0.35, and 0.4, the optical powers are -3, -1.25, 0.85, and 2.2 dBm. It is worth noting that the actual optical power compensation is lower than the theoretical DC responsivity reduction penalty. This is because the equalized response provides a higher relative weighting of the high-frequency components that govern signal transitions and bandwidth-induced ISI.

When $m$ is zero (no equalization), the bandwidth remains severely constrained by $RC$, yielding an open eye only at 30 Gbaud and increasingly blurred eye patterns at higher baud rates because of strong ISI. In contrast, open eye diagrams at 80, 90, and 100 Gbaud are obtained without DSP for $m$ = 0.2, 0.35, and 0.4, with corresponding SNR values of 3.62, 4.38, and 3.78. The 100-Gbaud limit observed at $m$ =0.4 is mainly determined by the aggregate bandwidth of the experimental setup, including the amplifier, Mach–Zehnder modulator, and RF probes. The measured 100-Gbaud eye diagram confirms the improvement of waveform integrity enabled by the selected equalization condition.

## 4. Discussion and conclusion

Table 1 provides a performance comparison of Ge PDs. The proposed EqPD successfully extends the measured 3-dB bandwidth of the vertical Ge PD beyond the 110-GHz measurement limit, representing, to the best of our knowledge, the highest reported bandwidth for a vertical Ge PD. Compared with conventional bandwidth-enhancement approaches based on device structure or parasitic optimization, the proposed frequency-domain equalization introduces a new approach to bandwidth extension by tailoring the differential frequency responses of two PD branches. Although the responsivity is reduced due to the differential subtraction mechanism, it can be further improved through device and equalization optimization.

Furthermore, compared with established equalization techniques such as feed-forward equalization (FFE), the EqPD enables frequency-response shaping directly at the photodetection stage, thereby alleviating the dependence on additional high-speed electronic equalization circuitry and mitigating the associated circuit-complexity and power-consumption overhead. In addition, the EqPD is inherently compatible with subsequent FFE, allowing both approaches to be jointly employed when further equalization is required. The EqPD is particularly suitable for high-speed systems with sufficient received optical power budget but whose performance is limited by PD bandwidth, where insufficient PD bandwidth can induce severe inter-symbol interference that cannot be mitigated simply by increasing optical power. Moreover, the EqPD is not limited to the Ge/Si platform and can be adapted to other PD structures that provide two branches with different frequency responses and differential readout. Furthermore, although a larger $m$ can give a higher 3-dB bandwidth, it does not necessarily yield the optimal link performance, as responsivity, bandwidth, and frequency-response flatness must be jointly considered, particularly when pronounced peaking occurs.

In principle, the bandwidth of EqPD can be further enhanced. The selected $PD_A$/$PD_B$ dimensions provide a design point that experimentally demonstrates the effectiveness of the EqPD, while further optimization of the PD geometries, such as reducing the Ge size to lower junction capacitance and increasing doping concentration of silicon region to reduce series resistance, could enable higher bandwidth enhancement for $PD_A$ and $PD_B$. The achievable bandwidth of the EqPD is strongly affected by the bandwidth of PDA. In addition, for systems with a predetermined bandwidth requirement, the tunable MZI can be replaced by a fixed-ratio multimode interferometer (MMI) to further simplify the implementation.

In conclusion, this work proposes a novel Ge PD that utilizes the frequency response of the high-bandwidth $PD_A$ to subtract the frequency response of the low-bandwidth $PD_B$. The frequency response of $PD_B$ attenuates more severely than that of $PD_A$ at high frequency. Experimental results show that the differential EqPD can extend the 3-dB bandwidth to beyond 110 GHz. A 100-Gbaud NRZ eye diagram has been demonstrated. It should be noted that the appropriate photocurrent ratio of $PD_A$ and $PD_B$ requires consideration of the trade-off between responsivity and bandwidth. This work establishes an effective pathway toward realizing high-performance PDs, offering high potential for the next generation of silicon photonic integrated circuits.

## Declaration of competing interest

The authors declare that they have no known competing financial interests or personal relationships that could have appeared to influence the work reported in this paper.